\documentclass[12pt]{article}
\usepackage[lmargin=.75in,rmargin=.75in,tmargin=1.in,bmargin=1in]{geometry}

\usepackage{graphicx, epsfig}
\usepackage{color}
\usepackage{mathrsfs}

\newcommand{\be}{\begin{equation}}
\newcommand{\ee}{\end{equation}}
\newcommand{\bea}{\begin{eqnarray}}
\newcommand{\eea}{\end{eqnarray}}

\newcommand{\gapp}{\mathrel{\raise.3ex\hbox{$>$}\mkern-14mu
              \lower0.6ex\hbox{$\sim$}}}
\newcommand{\lapp}{\mathrel{\raise.3ex\hbox{$<$}\mkern-14mu
              \lower0.6ex\hbox{$\sim$}}}

\begin{document}
\centerline{Essay written for the Gravity Research Foundation 2013 Awards for Essays on Gravitation.}
\centerline{March 20, 2013}

\vspace{1cm}

\centerline{\large \bf Implications of the Higgs discovery for gravity and cosmology}
\smallskip
\centerline{Dejan Stojkovic}
{\it HEPCOS, Department of Physics, SUNY at Buffalo, Buffalo, NY 14260-1500 }

\centerline{ email: {\it ds77@buffalo.edu}}


\begin{abstract}
\noindent
The discovery of the Higgs boson is one of the greatest discoveries in this century. The standard model is finally complete.
Apart from its significance in particle physics, this discovery has profound implications for gravity and cosmology in particular.
Many perturbative quantum gravity interactions involving scalars are not suppressed by powers of Planck mass. Since gravity couples anything with mass to anything with mass, then Higgs must be strongly coupled to any other fundamental scalar in nature, even if the gauge couplings are absent in the original Lagrangian. Since the LHC data indicate that the Higgs is very much standard model-like, there is very little room for non-standard model processes, e.g. invisible decays. This severely complicates any model that involves light enough scalar that the Higgs can kinematically decay to. Most notably, these are the quintessence models, models including light axions, and light scalar dark matter models.
\end{abstract}

Quantum gravity is one of the most elusive concepts in modern physics. Despite concerted efforts of many researchers over many decades, it appears that we are still far from the satisfactory self-consistent theory of quantum gravity. This is the reason why there are so few (if any) robust predictions of quantum gravity effects. Anyway, we still have some ideas of what quantum gravity should and should not look like. The standard lore is that that quantum gravity should decouple from the rest of physics at low energies, or otherwise we would have noticed strange phenomena which can not be explained by the low energy effective theories like the standard model of particle physics. This expectation is embodied in the requirement that dangerous quantum gravity effect must be suppressed by powers of the scale of quantum gravity, i.e. the Planck mass $M_{\rm Pl}$. Indeed, in four space-time dimensions, for an action to be dimensionless, Lagrangian density must have the mass dimension of four. Any operator with mass dimension greater than four must be therefore suppressed by powers of the relevant physical scale. If the relevant physics is quantum gravity, then the suppression scale is  $M_{\rm Pl}$. For example, one of the potentially dangerous operators is the four-fermion operator
\be \label{ff}
\frac{1}{M^2_{\rm Pl}} \psi \psi \psi \psi .
\ee
Since the mass dimension of the fermion operator is $3/2$, the four-fermion operator must be suppressed by two powers of $M_{\rm Pl}$. This operator is dangerous since it could mediate gravitationally induced proton decay. For example, an interaction involving three quarks and one lepton, i.e. QQQL, is all it takes for a proton to decay. In the standard model this interaction mediated by gauge couplings is suppressed since it inevitably involves violation of the baryon number. The standard model allows for the baryon number violation (which is a crucial ingredient in the standard model baryogenesis) but only on the non-perturbative level which involves instantons. Such processes, which may have played important role in early universe, are thus highly suppressed at temperatures below the electroweak symmetry breaking temperatures. However, conservation of baryon number is just an accidental global low-energy effective symmetry, and quantum gravity is not expected to preserve it (as opposed to local gauge symmetries which are dynamically preserved) \cite{Stojkovic:2005zq}.  We see from Eq.~(\ref{ff}) that perturbative gravity effects can still mediate proton decay. Fortunately for us, the operator in Eq.~(\ref{ff}) yields the proton life-time of
\be
\tau_{\rm proton} \approx m^{-1}_{\rm proton} \left(\frac{M_{\rm Pl}}{m_{\rm proton}}  \right)^4
\ee
which for $M_{\rm Pl} = 10^{19}$GeV is of the order of $10^{45}$ years, and thus safely above any current proton life-time limits ($\sim 10^{34}$years). Another potentially dangerous quantum gravity induced process comes from neutron-antineutron oscillations. An operator responsible for this process is ``UDDUDD" where ``U" stands for the up-quark and ``D" for the down quark. However, this is a dimension-nine operator, so it is suppressed by five powers of $M_{\rm Pl}$, which lifts this process safely above any current experimental limits. Once again, $M_{\rm Pl}$ suppression decouples quantum gravity processes from low energy phenomenology.

However, the real danger comes in the form of the scalar fields. Note that a four-scalar interaction is only a dimension four operator since scalars have the mass dimension one.
Thus operators of the sort
\be \label{p}
\phi \phi \phi \phi \, , \ \ \ \ \ \ \  \phi \phi \chi \chi \, \ \ \ \ \ \ \  {\ldots}
\ee
where $\phi$ and $\chi$ are some fundamental scalar fields, are not suppressed by any powers of $M_{\rm Pl}$.
It is also possible to write other operators like
\be \label{np}
M_{\rm Pl} \phi \phi \phi \, , \ \ \ \ \ \ \   M_{\rm Pl} \phi \chi \chi \, , \ \ \ \ \ \ \ M^2_{\rm Pl} \phi \phi  \, , \ \ \ {\ldots}
\ee
but they would represent non-perturbative effects. In the absence of the fully fledged theory of quantum gravity, it is much more uncertain what happens with the non-perturbative effects. In particular, it may (or may not) happen that that they will suffer from a large action suppression in the form of $e^{-S}$. An action, $S$, for non-perturbative processes might involve exotic topology changing (wormhole) effects. The value of this action might be large, thus suppressing such processes. However, the existence of wormholes and the value of their action are very sensitive to the structure of space near  $M_{\rm Pl}$, and it is thus connected with large uncertainties \cite{Kallosh:1995hi}.

On the other side, we believe that perturbative gravitational effects are much more under control, so we will concentrate here on them.
Of course, there is always a possibility that even perturbative processes described by dimension-four operators like in Eq.~(\ref{p}) are suppressed by some unknown factors.
However, there exist explicit calculations that indicate this is not the case. In \cite{Hawking:1995ag}, Hawking  derived the general form of the scattering amplitude for the processes involving particles of different spins using the path integral approach to quantum gravity. In particular, he managed to include virtual quantum gravity effects in the scattering processes by integrating over the space-time metrics including virtual black holes consistent with the symmetries. The result in a compact form is
\be \label{h}
M^{4-2n(1+s)}_{\rm Pl} \phi^{2n}
\ee
where $s$ is the spin of the particle and $n \geq 2$ is the number of pairs of ingoing or outgoing  scattered momenta. For scalar particles with $s=0$, and $n=2$, we recover the expression in Eq.~(\ref{p}). It is clear that gravitationally induced scalar-scalar interaction of dimension four is not forbidden nor suppressed by any small factors.

One of the most important implications of the result in Eq.~(\ref{h}) is that one can not avoid a strong scalar-scalar interaction even if it is absent in the original Lagrangian.
In other words any scalar field in nature should be coupled to any other scalar field with the coefficient of the order of one.

While fundamental scalar fields are ubiquitous in modern physics, until recently we had no evidence that any of them actually existed. Then, in 2012, the Large Hadron Collider at CERN made a historic discovery of the Higgs boson. The standard model is now complete. However, the fact that is not emphasized enough is that the Higgs is the first known fundamental scalar in nature. For the first time in history, we can now make some important statements that we were not able to make before. In particular, we know the mass of the Higgs, which is around $m_H = 126$GeV, so we definitely know which scalar-scalar processes are kinematically allowed. Since both ATLAS and CMS data indicate that the Higgs is very much standard-model-like, there is very little room for non-standard model processes, e.g. invisible decays \cite{Djouadi:2012zc,Djouadi:2011aa}. In near future, we expect this room to shrink even more.

Implications of the Higgs discovery for our understanding of the Universe might be profound. Two of the most important problems in cosmology are certainly the dark energy and dark matter problems. In order to solve the dark energy problem, many models postulating the existence of a new fundamental scalar field were put forward. Not making a bookkeeping distinction among them, we will call the whole class of cosmological models where a slowly rolling scalar field dominated the late-time energy density of the universe and drives accelerated expansion of the universe simply quintessence models. For most of the quintessence models to work, the Compton wavelength of this scalar field must be of the order of the current Hubble radius. This implies that its  mass must be extremely small, of the order of $m_q = 10^{-33}$eV. Such an extremely light particle is very difficult to avoid detection. For this reason it is usually postulated that the coupling of this field to the rest of the standard model fields is highly suppressed. The gauge couplings can indeed be eliminated by requiring that the quintessence field is a gauge singlet. However, gravity couples everything with mass to everything with mass, so couplings as in Eq.~(\ref{p}) can not be avoided. Since we now know that the Higgs exists, and according to Eq.~(\ref{p}) it interacts strongly with the quintessence field, we also know that the Higgs can kinematically decay into the quintessence field quanta, for example as $\phi \rightarrow 2 \chi$. [Note that the coupling $\phi \phi \chi \chi$ will yield a tree level vertex $\phi_{\rm ph} \chi \chi$ between the physical Higgs, $\phi_{\rm ph}$, and the scalar field $\chi$ after the symmetry breaking when we expand around the vacuum $v$ as $\phi = v + \phi_{\rm ph}$.]
Since this process does not violate any symmetry (gauge or global), this decay should proceed essentially unsuppressed. However, this would be in violation of the LHC experimental data, from which we know that the disagreement with the standard model decay rates can not be large. Thus, the discovery of the Higgs  severely complicates the validity of the quintessence models. The light axion models are in the same danger. It seems very implausible that such very light scalar fields can exist in nature.

A similar argument can be applied to the dark matter problem. In order to explain the missing mass in galaxies and galaxy clusters we have to postulate the existence of new particles which interact very weakly with the standard model particles. While dark matter with mass around a TeV is preferred for some theoretical reasons, it has been argued that the tension between different direct searches for dark matter can be relieved by light dark matter with the mass of a few GeV (see e.g. \cite{Hooper:2012cw}). If these particles are scalars, from our previous discussion we know that the Higgs is kinematically allowed to decay into them, which would in turn be in violation of what we know about the Higgs from the collider data\footnote{At this moment there is still some room for invisible Higgs decays into other scalar particles. We expect this window to close in very near future.}. Thus, the sole existence of the Higgs with the given mass has strong implications for the existence of the light scalar dark matter. In particular, a scalar dark matter particle can not be lighter than $m_H/2 = 63$GeV.

As a good cross-check, from Eq.~(\ref{p}) we see that the Higgs quartic self-interaction ($\lambda \phi^4$) must be unsuppressed too. In other words, the Higgs quartic coupling constant $\lambda$ must be of the order of unity. Indeed, the Higgs vacuum expectation value $v = 246$GeV and the Higgs mass $m_H = \lambda v = 126$GeV indicate that $\lambda$ is of the order of unity. On the other hand, Yukawa-like coupling of the Higgs with fermions $\phi \bar{\psi} \psi$ is also a dimension-four operator, but is not of the form of the four-scalar interaction as in Eq.~(\ref{h}), so strictly speaking this amplitude is unknown. One of the operators that is allowed by Eq.~(\ref{p}) is $(1/M_{\rm Pl}) \phi \phi \psi \psi$, but since it is suppressed by one power of $M_{\rm Pl}$ it is not clear if it represents an obvious danger to potential interaction of the Higgs with fermions.

Phenomenology of the inflation field, which drove primordial inflation in the early universe, is not affected by the existence of the Higgs. It is true that the inflation field could quickly decay into the Higgs field, but that is an expected feature of the inflaton anyway.

In conclusion,  the sole knowledge that a fundamental scalar field exists, along with the value of its mass, puts strong constraints on all the other  models that also include fundamental scalar fields. The reason is unsuppressed perturbative gravitational coupling of the scalar fields, which is expected from the simple dimensional analysis but it is also backed up with explicit loop calculations which include perturbative quantum gravity effects \cite{Hawking:1995ag}. The opposite statement is also very interesting - if we ever discover another light scalar field, this would invalidate Hawking's calculations in \cite{Hawking:1995ag} and cast a serious shadow on the Euclidean path integral approach to quantum gravity.

\bigskip

\centerline{Acknowledgments}
DS acknowledges the financial support from NSF, grant number PHY-1066278.

\end{document}